\documentclass[aps,prl,superscriptaddress,twocolumn,longbibliography]{revtex4-1}
\usepackage{bm} \usepackage{graphicx} \usepackage{amsmath}
\usepackage{amsthm,stmaryrd}
\usepackage{amssymb} \usepackage{epstopdf} \usepackage{txfonts}
\usepackage{color}
\usepackage{ytableau}
\usepackage{verbatim} 
\usepackage{soul} 

\newcommand{\ket}[1]{|{#1}\rangle}
\newcommand{\bra}[1]{\langle{#1}|}
\newcommand{\mean}[1]{\langle{#1}\rangle}

\DeclareRobustCommand\openzero{\leavevmode\hbox{0\kern-.55em0}}
\mathchardef\minus="002D
\newtheorem{thm}{Theorem}
\newtheorem*{cor}{Corollary}

\newcommand{\mysection}[1]{{\it #1}-- }

\begin{document}

\title{
  Multiphoton Tomography with Linear Optics and Photon Counting
}

\author{ Leonardo Banchi}
\affiliation{QOLS, Blackett Laboratory, Imperial College London, London SW7 2AZ, United Kingdom}

\author{ W. Steven Kolthammer }
\affiliation{QOLS, Blackett Laboratory, Imperial College London, London SW7 2AZ, United Kingdom}

\author{ M.S. Kim}
\affiliation{QOLS, Blackett Laboratory, Imperial College London, London SW7 2AZ, United Kingdom}

\date{\today}
\begin{abstract}
	Determining an unknown quantum state from an ensemble of identical systems is a fundamental, yet experimentally demanding, task in quantum science. Here we study the number of measurement bases needed to fully characterize an arbitrary multi-mode state containing a definite number of photons, or an arbitrary mixture of such states. We show this task can be achieved using only linear optics and photon counting, which yield a practical though non-universal set of projective measurements. We derive the minimum number of measurement settings required and numerically show that this lower bound is saturated with random linear optics configurations, such as when the corresponding unitary transformation is  Haar-random. Furthermore, we show that for $N$ photons, any unitary $2N$-design can be used to derive an analytical, though non-optimal, state reconstruction protocol. 
\end{abstract}

\maketitle

\mysection{Introduction}
An unknown quantum state can be determined by making a set of suitable measurements on identically prepared copies \cite{d2003quantum,james2001measurement,banaszek1999maximum,christandl2012reliable,paul1996photon}. This procedure, known as quantum state tomography, is a fundamental concept in quantum science with wide ranging applications. For example, tomography allows one to assess quantum systems for use in quantum information processing by quantifying resources such as entanglement
\cite{horodecki2009quantum}, quantum correlations \cite{modi2010unified}, and coherence \cite{streltsov2017colloquium}. Indeed, since most measures of these resources require complete knowledge of the density matrix describing a system, full quantum tomography is often necessary. Similarly, tomography can be applied to quantum sensing \cite{degen2017quantum,braun2017quantum} to evaluate the capacity of a quantum probe state to yield enhanced measurement precision \cite{giovannetti2011advances,vidrighin2014joint}.

A well-established framework for photonic quantum information uses a single
photon and multiple modes to encode discrete-variable quantum states. A qubit
may be encoded using a single-photon, two-mode state \cite{chuang1995Simple}, and a qudit may be encoded by incorporating additional modes \cite{langford2004measuring}.
Multi-qubit states of this form have been employed widely, including entanglement-based
quantum-key distribution \cite{gisin2002quantum}, quantum simulation \cite{pitsios2017photonic}, tests of quantum nonlocality
\cite{brunner2014bell}, entanglement generation \cite{wang2018multidimensional},
and linear optical quantum computing \cite{kok2007linear}. 
For these states, optical tomography can be
readily achieved using combinations of single-qudit measurements
\cite{james2001measurement,thew2002qudit}, which require only linear optics and
single-photon detection. Exact reconstruction of $N$ qubits can thus be achieved using $2^N+1$ measurement bases.  Using this method, full tomography of up
to six single-photon qubits has been demonstrated
\cite{schwemmer2014experimental}.

However, this approach to optical tomography does not apply to more general states of multiple modes containing a definite total number of photons. In this case, a mode may contain multiple photons, which enables new applications including approaches to quantum sampling  \cite{aaronson2011computational}, imaging \cite{humphreys2013quantum}, and error-correction \cite{chuang1997bosonic,wasilewski2007protecting}. 
An alternate approach to state tomography for such states is to use balanced homodyne detection and well-developed continuous-variable algorithms to reconstruct the phase-space Wigner function \cite{smithey1993measurement, d1999universal,lvovsky2009continuous}. In the general continuous variable setting, however, only partial reconstruction is possible with a finite number of measurement settings. Furthermore, this detection scheme adds substantial experimental requirements, including access to a mode-matched, multimode phase-stable local oscillator. In contrast, since the state has a definite photon number, tomographically complete measurements can theoretically be formulated using a finite number of measurement bases. Whether or not these measurement bases can be achieved using photon counting, though, has not been previously known.

Here we prove that an arbitrary state of $N$ indistinguishable photons in $M$ modes can be reconstructed using a finite number of measurement bases that correspond to different configurations of an $M$-mode linear-optical interferometer followed by photon counting. Notably, this result is not limited to states that can be created from Fock states using linear optics.  Furthermore, we derive a minimal number of interferometer configurations required for a given $N$ and $M$.

Our results extend to arbitrary mixtures of states with fixed, but possibly different, number of photons and to measurement strategies that incorporate additional modes through the use of ancillary vacuum states. As the number of measured modes increases, the required number of interferometer configurations decreases, eventually reaching one. In this limit, our work relates to previous studies of tomography using a single measurement basis in an extended Hilbert space \cite{dariano2002universal,allahverdyan2004determining}, a concept first applied experimentally to nuclear spins \cite{du2006realization} and then to single-photon qubits measured using a multimode quantum walk \cite{zhao2015experimental,bian2015realization}. The latter approach was recently extended to two-photon, two-mode states using a six-mode interferometer and it was conjectured this method would work for larger systems \cite{titchener2016two,titchener2017scalable}. Related work has investigated how the number of additional modes required for high-fidelity state estimation depends on the purity of the input state \cite{oren2017quantum}. Our results generalize these photonic studies that use a single measurement configuration by proving tomographic feasibility, deriving a bound on the minimum number of measurement modes, and providing an explicit reconstruction protocol. 

We numerically show that use of random interferometer configurations, in particular those corresponding to Haar-random transformations, enable tomography using the minimum number of configurations. Additionally, we derive an analytical algorithm for state tomography that employs any unitary $2N$-design \cite{roy2009unitary}, thus generalising a known result for qudit systems \cite{roy2007weighted} to the multi-photon case. While unitary designs are not optimal for our task, an advantage is they have been 
extensively studied in the past for their relevance in many quantum information 
theory protocols \cite{ambainis2007quantum} and quantum metrology \cite{oszmaniec2016random}. Indeed, unitary
designs can be obtained either with random circuits 
\cite{hayden2004randomizing,brandao2016local,dankert2009exact,brown2010convergence}, 
random basis switching \cite{nakata2017efficient} or, more physically, by applying 
random pulses to a controllable system \cite{banchi2017driven}.


\begin{figure}[t]
	\centering
	\includegraphics[width=\linewidth]{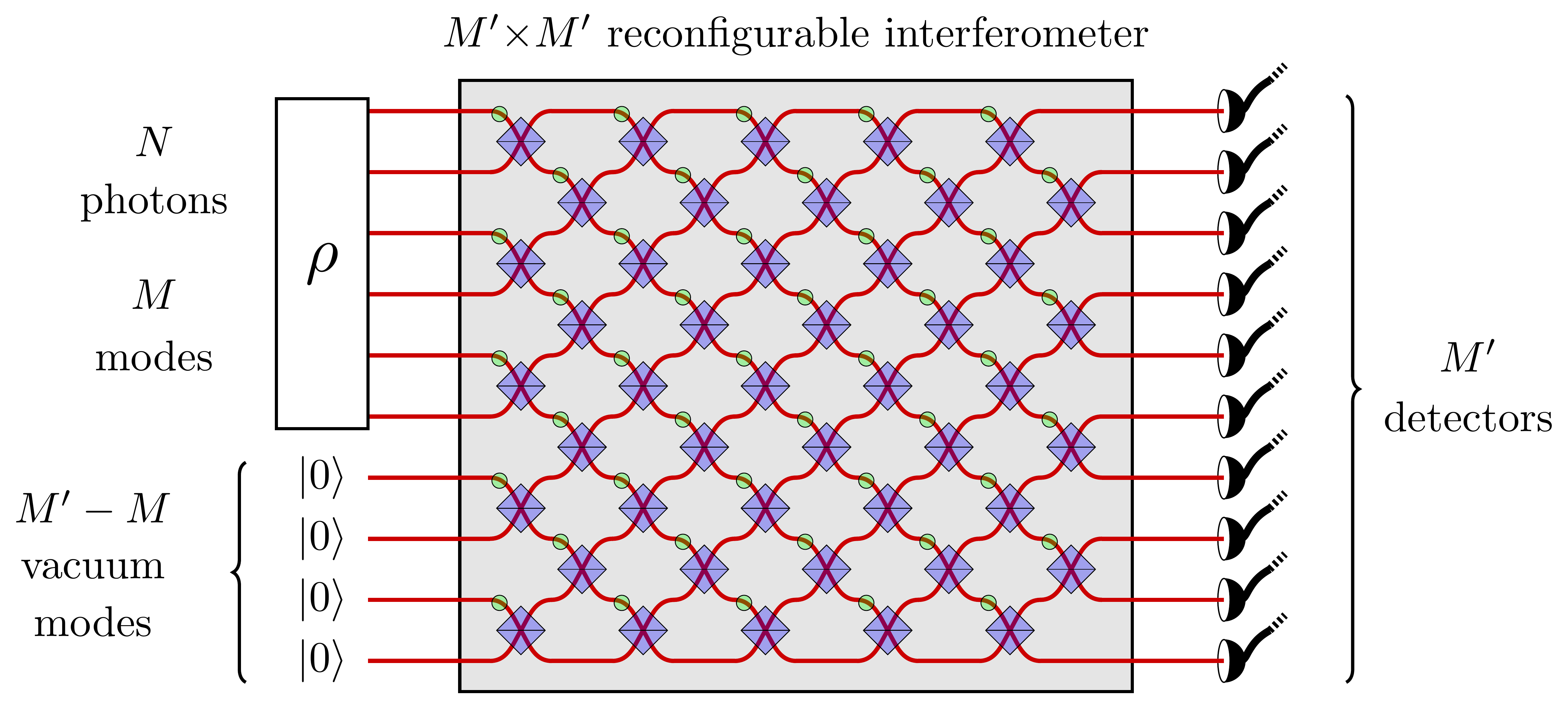}
	\caption{Tomography of a generic unknown state 
		$\rho$ of $N$ photons in $M$ modes. 
		Our protocol uses configurations of an $M'$-mode linear optical interferometer followed by photon counting. When $M'>M$, 
		vacuum modes are appended to the state $\rho$. 
	}
	\label{fig:linearnet2}
\end{figure}

\mysection{Feasibility of tomography}
Consider a generic quantum state of $N$ indistinguishable photons in $M$ modes.
Our goal is to completely characterize the state by measuring multiple copies
of it using linear optics and photon counting, as illustrated in Fig.~\ref{fig:linearnet2}.
In this approach, a measurement basis corresponds to a particular configuration
of linear optics. We also allow for measurements over $M' \ge M$ modes,
achieved by appending $M'-M$ vacuum modes to the state of interest. Our first
main result is that full tomography can always be achieved using a finite
number of measurement configurations:
\begin{thm} \label{thm:feasbility}
	An $N$-photon, $M$-mode state can be reconstructed using photon counting and $M$-mode linear optical interferometer with a finite number $R$ of configurations, where
	 \begin{equation}
	 R < \binom{N+M^2-1}{N}^2.
	 \end{equation}
\end{thm}

The theorem is proved by building an explicit reconstruction algorithm.
Let $\ket{\nu}$ be the multi-mode Fock basis 
$\ket{\nu}\equiv \ket{k_1,\dots,k_{M}}$, where $k_j$ is the number of particles in mode $j$ and 
$\sum_j k_j = N$, while we 
use a prime to denote a Fock basis 
$\ket{\nu'}\equiv \ket{k_1,\dots,k_{M'}}$, where the number of output modes 
$M'$ may be higher than the number of inputs $M$. 
Moreover, let $U(g)$ be a set of available unitary operations that can be made in the system. 
In linear optics the most general SU(M$'$) transformation can be obtained with a 
collection of beam splitters and phase shifters \cite{clements2016optimal}, as shown in Fig.~\ref{fig:linearnet2}. 
Such transformation can be expressed in the second quantized notation as 
$U(g)=e^{i \sum_{kl} H_{kl} a_k^\dagger a_l}$, 
where $g=e^{i H}$ is a $M'\times M'$ unitary matrix. 

State tomography requires reconstruction of the state $\rho$ from measurement outcomes, each specified by a series of photon counts $\nu'$. These outcome probabilities are readily calculated as $p_{\nu',g} = \bra{\nu'} U(g)^\dagger \rho U(g)\ket{\nu'}$ for a specified interferometer configuration $g$. Expanding the above equation gives
\begin{equation}
	p_{\nu',g} = \sum_{\alpha,\beta} \bra{\nu'} U(g)^\dagger \ket\alpha\bra\alpha\rho
	\ket\beta\bra\beta U(g)\ket{\nu'} \equiv [\mathcal L(\rho)]_{\nu',g}~,
	\label{eq:pfromrho}
\end{equation}
with the superoperator $\mathcal L_{\nu' g,\alpha\beta} = 
\bra{\nu'} U(g)^\dagger \ket\alpha\bra\beta U(g)\ket{\nu'} $. 
The superoperator $\mathcal L$ is constructed using different configurations $g_j$, with $j=1,\dots,R$. 
The numbers $\alpha$ and $\beta$ index the elements of the Fock space, whose dimension is 
$D_{N,M} = \binom{N+M-1}{N}$, while $\nu'=1,\dots,D_{N,M'}$. 
As such, $\mathcal L$ is normally a rectangular operator. 
Tomography is possible if there is a large enough $R$ such that the 
linear system \eqref{eq:pfromrho} admits a unique solution for any $p$. 
A unique solution is obtained \cite{klose2001measuring} when the Gramian matrix 
$\mathcal L^\dag\mathcal L$ has full rank. 
%
In this case, the best reconstruction algorithm \cite{klose2001measuring}
is given by the pseudo-inverse 
$\rho_{\rm best} := (\mathcal L^\dagger \mathcal L)^{-1} \mathcal L^\dag[p]$, 
which is always the best fit solution that minimizes 
the least-square error.

For any linear optics configuration $g$, the matrix elements 
$\bra\beta U(g)\ket{\nu'}$ can be calculated exactly, either using 
combinatorial expressions or matrix permanents 
\cite{scheel2004permanents,biedenharn1985u,aaronson2011computational}: 
for $\ket{\alpha} = \ket{a_1,a_2,\dots}$ and
$\ket{\beta} = \ket{b_1,b_2,\dots}$, one finds 
$
	\bra{\alpha}U(g)\ket{\beta} = 
	{{\rm per}(g_{\{\alpha,\beta\}})}{/}{\sqrt{\alpha!\beta!}}
$
where $\alpha!=a_1!a_2!\dots$, and  similarly for $\beta!$, while
$	g_{\{\alpha,\beta\}} $ is the $N\times N$ matrix 
obtained by copying $a_i$ times 
the $i$-th columns of $g$ and, and $b_j$ times the $j$-th row of $g$. 
Although the computation of the matrix permanent 
is \#P-hard, it is still possible for the values of $N$ and $M$ available in 
near-term devices \cite{neville2017classical}. 
Moreover, there are cases for which specific values of the permanent 
can be computed analytically 
\cite{tichy2014stringent,dittel2017many,viggianiello2018experimental}.
In the worst case, without making any simplifications about the permanents, 
in the Supplementary Material we show that the number of operations to reconstruct the state 
from Eq.~\eqref{eq:pfromrho} is 
$\mathcal O({\rm poly}(D_{N,M}, 2^N))$. Therefore, as in qubit systems, the 
difficulty is mostly due to exponentially growing Hilbert space, rather than 
to the complexity of the permanent. 

Given the above framework, we now sketch our proof of Theorem~\ref{thm:feasbility}, which is elaborated in the Supplementary Material. In particular, we show that with interferometer configurations $\{g_j\}_{j=1,\dots,R}$ corresponding to a unitary $2N$-design, exact reconstruction is possible from experimental measurements of $p_{\nu',g_j}$ for all $j=1,\dots,R$. Our theorem then follows from known properties of unitary designs \cite{roy2009unitary}: they exist for all $N$ and $M$, and their size is bounded by $R < D_{N,M^2}^2$.

To connect our tomographic task to unitary designs, we first note that the matrix $\mathcal L$ 
is composed by $U(g)\otimes U(g)^*$ matrices. Although $U(g)$ is an irreducible 
representation of $g$, $U(g)\otimes U(g)^*$ is not, and indeed it can be written 
as a direct sum of Wigner-$D$ matrices $\mathcal D^{\lambda_r}_{m,m'}$ where 
$\lambda_r$ refer to different irreducible representations and $m,m'$ are 
Gelfand-Tsetlin patterns that index the different states 
(see Supplementary Material).
Since the matrices $\mathcal D^{\lambda_r}_{m,m'}(g)$ are orthogonal over $g$ and $\mathcal L\propto \mathcal D(g)$, one can use the matrix 
$\mathcal D(g)^* $ 
to construct an operator $X^{\nu'}_{\alpha\beta}(g)$ such that 
$\bra\alpha\rho\ket\beta = \sum_{\nu'} \int dg\, X^{\nu'}_{\alpha\beta}(g)  p_{\nu',g}$,
where $p_{\nu',g}$ are the outcome probabilities in Eq.~\eqref{eq:pfromrho}.

Tomography is therefore achieved via a formal average over the continuous group. However, this is not practical as it would require an infinite number of measurement configurations. 
Instead we use the theory of weighted unitary designs \cite{roy2009unitary}, to replace the continuous average with a discrete average over a discrete set of unitaries $g_j$. A $q$-design is a discrete set of unitaries such that the weighted average 
of group functions $f(g)$ over those unitaries is equal to the average 
over the continuous group $\int dg f(g)$, provided that $f(g)$ is a polynomial 
of at most degree $q$ in $g$ and $g^*$. 
Since the matrices $\mathcal D(g)$ are a polynomial of at most 
degree $N$ in $g$ and $g^*$, one can choose any weighted $2N$-design protocol 
to analytically perform full-state tomography, as shown in the Supplementary Material. 
Calling $g_j$ those unitaries, 
$\bra\alpha\rho\ket\beta = \sum_{\nu',j} X^{\nu'}_{\alpha\beta}(g_j)  p_{\nu',g_j}$. 
This concludes the proof of Theorem~\ref{thm:feasbility}. We note however that 
unitary $2N$-designs satisfy a more stringent requirement than the simpler
inversion of Eq.~\eqref{eq:pfromrho}, and consequently, this approach is generally not optimal in terms of the number of measurement configurations used. 
Theorem~\ref{thm:feasbility} can be trivially extended to mixtures 
$\rho=\sum_{N=1}^{N_{\rm max}} \pi_N\,\rho_N$ where $\rho_N$ is a $M$-mode
$N$-photon state, as each $N$-photon state can be reconstructed independently
via postselection (see Supplementary Material).

\mysection{Minimum measurement configurations}
We now consider the minimum number of linear optics configurations $R$ required to achieve tomography. Our second main result gives a lower bound on the number of configurations required:
\begin{thm}\label{thm:count}
	An $N$-photon, $M$-mode state can be reconstructed with photon counting and an $M$-mode linear optical interferometer using at least 
	\begin{equation}
		R_{N,M} = \binom{N + M}{N} - \binom{N + M - 2}{M}
			\label{count}
	\end{equation}
	configurations. More generally, for an interferometer with $M'>M\ge 2$ modes and ancillary vacuum states, the minimal number of reconfigurations is 
	\begin{equation}
		R_{N,M,M'} = \left[\frac{ (N+M-2)!  (M'-2)! }{(N+M'-2)! (M-2)! } R_{N,M}\right]~,
		\label{countp}
	\end{equation}
	where $[x]$ is the smallest integer greatest or equal to $x$. 
\end{thm}
Equation~\eqref{count} shows that the number of measurement configurations is larger than estimated from a simple counting argument. In particular, the number of $M$-mode Fock states with $N$ total photons, $D_{N,M}=\binom{N+M-1}{N}$, gives the dimension of the symmetric Hilbert space. A generic state is thus specified by $D_{N,M}^2-1$ independent elements.

A single measurement configuration involves $D_{N,M}$ different outcomes, which provide $D_{N,M}-1$ independent parameters. Therefore, one may expect that $D_{N,M}+1$ configuration may be sufficient for full state reconstruction. Instead, our theorem shows a larger number is required, $R_{N,M} >D_{N,M}+1$. This increased requirement is due to linear optics providing only a subset of the possible unitary operations on the multi-particle state.  
Nonetheless, complete tomography with a smaller set of 
configurations is possible with ancillary output modes, 
as $R_{N,M,M'} < D_{N,M}+1 < R_{N,M}$ for any $M'>M$.

For the two-mode case, $M=2$, an explicit measurement protocol which saturates our bound $R_{N,2}=2N+1$ is known \cite{walser1997measuring}. This protocol exploits the Schwinger boson formalism that maps our problem onto the tomography of a spin $S=N/2$, allowing the use of known algorithms for large spin systems \cite{newton1968measurability,klose2001measuring,hofmann2004quantum}. However, this approach exploits properties of SU(2) representations that cannot be easily adapted to larger $M$ \cite{filippov2009spin,tan20133}.
Our theorem generalizes the above construction to the general multi-mode case.

Two proofs of Theorem~\ref{thm:count} are presented in the Supplementary Material, one 
based on representation theory and one based on irreducible tensors. Here 
we briefly describe the main steps of the second proof. Measuring diagonal elements 
in the Fock basis is equivalent to the measurement of all the expectation values 
of polynomials of number operators $T^k_k = a^\dagger_ka_k$. 
According to Wick's theorem,
all independent polynomials in the number operators can be written via the rank $r$ tensors 
$T^{k_1,\dots,k_r}_{k_1,\dots,k_r} = 
a^\dagger_{k_1}\cdots a^\dagger_{k_r} a_{k_1}\cdots a_{k_r}$. However, not all 
$\mean{T^{k_1,\dots,k_r}_{k_1,\dots,k_r}}$ are independent. 
For instance, if one measures $\mean{T^k_k}$ for $k=1,\dots,M-1$, then one gets 
$\mean{T^M_M}=N-\sum_{k=1}^{M-1}\mean{T_k^k}$ without further measurements. 
In the Supplementary Material we show that the number of independent rank-$r$ tensors 
is $D_{r,M-1}$. Their expectation value for $r=1,\dots,N$ completely and uniquely 
specify photodetection measurements. 
Similarly, the full state is completely and uniquely 
specified by the expectation value of 
the tensors 
$ T^{k_1,\dots,k_r}_{\ell_1,\dots,\ell_r} = 
a^\dagger_{k_1}\cdots a^\dagger_{k_r} a_{\ell_1}\cdots a_{\ell_r}$. 
The number of such independent rank-$r$ tensors is $D_{r,M}^2-D_{r-1,M}^2$.

Tomography then consists in reconstructing the expectation value of 
off-diagonal tensors from the
measurement of $\mean{T^{k_1,\dots,k_r}_{k_1,\dots,k_r}}$ 
after different configurations $U(g)$.  Since the latter corresponds to 
$\mean{U(g)^\dagger T^{k_1,\dots,k_r}_{k_1,\dots,k_r}U(g)} = 
[g^{\dagger \otimes r} \mean{T} 
g^{\otimes r }]^{k_1,\dots,k_r}_{k_1,\dots,k_r}$,
all off-diagonal tensors with different rank $r$ can be 
reconstructed independently for $r=1,\dots,N$. The most difficult tensor to 
reconstruct is then that with $r=N$. Via dimensional counting, this 
reconstruction requires $[D_{N,M}^2-D_{N-1,M}^2]/D_{N,M-1} \equiv R_{N,M}$ 
transformations. Equation \eqref{count} follows by assuming that the same 
configurations are sufficient for reconstructing even lower rank tensor. 
This latter assumption is the reason why Eq.~\eqref{count} is a lower bound. 
Similarly, Eq.~\eqref{countp} appears for different number of modes 
as $R_{N,M,M'}=[(D_{N,M}^2-D_{N-1,M}^2)/D_{N,M'-1}]\equiv [R_{N,M} D_{N,M-1}/D_{N,M'-1}]$.

Theorem~\ref{thm:count} can be extended to mixtures 
$\rho=\sum_{N=1}^{N_{\rm max}} \pi_N\,\rho_N$ where $\rho_N$ is a $M$-mode
$N$-photon state. In this case, the minimal number of settings is 
$\max_{N\le N_{\rm max}}R_{N,M,M'}$ (see Supplementary Material). 
Theorem~\ref{thm:count} also determines the number of ancillary modes needed to achieve tomography with a single measurement configuration:
\begin{cor}
	An $N$-photon, $M$-mode state can be reconstructed with a single configuration of an $M'$-mode 
	linear optical interferometer if 
	\begin{equation}
		D_{N,M'-1} \ge D_{N,M}^2-D_{N-1,M}^2 = R_{N,M} D_{N,M-1}~.
		\label{corollary}
	\end{equation}
\end{cor}
The scaling of Eq.~\eqref{corollary} can be investigated for large $N$ and $M$ using the entropic expansion $ \binom{n}{k} \approx 2^{n H_2(k/n)}$, where $H_2(x) = {-}x\log_2x{-}(1{-}x)\log_2(1{-}x)$ is the binary entropy. 
If additionally $N\gg M$, we find that $R_{N,M} D_{N,M-1}\approx N^{2M-3}$, and 
$	D_{N,M'-1} \approx N^{M'-2}$. Therefore the minimum
number of measurement modes required is given by 
\begin{equation}
	M' \stackrel{N\gg M}{\gtrsim} 2M-1~.
	\label{onelimit}
\end{equation}
In this limit, tomography can be achieved using a single measurement configuration with photon counting over twice as many modes as the input state, and this result is independent of $N$.

In the opposite limit $N\ll M$, we approximate $\binom{N+M}{M}\approx M^N/N!$ to find 
\begin{equation}
	M' \stackrel{N\ll M}{\gtrsim} \frac{M^2}{\sqrt[N]{N!}}~.
	\label{onelimit2}
\end{equation}
This seemingly counterintuitive result shows that the required number of measured modes decreases as the number of photons increases. This is due to the large increase in number of measurement outcomes that results from an increase in the number of photons.

\begin{figure}[t]
	\centering
	\includegraphics[width=0.95\linewidth]{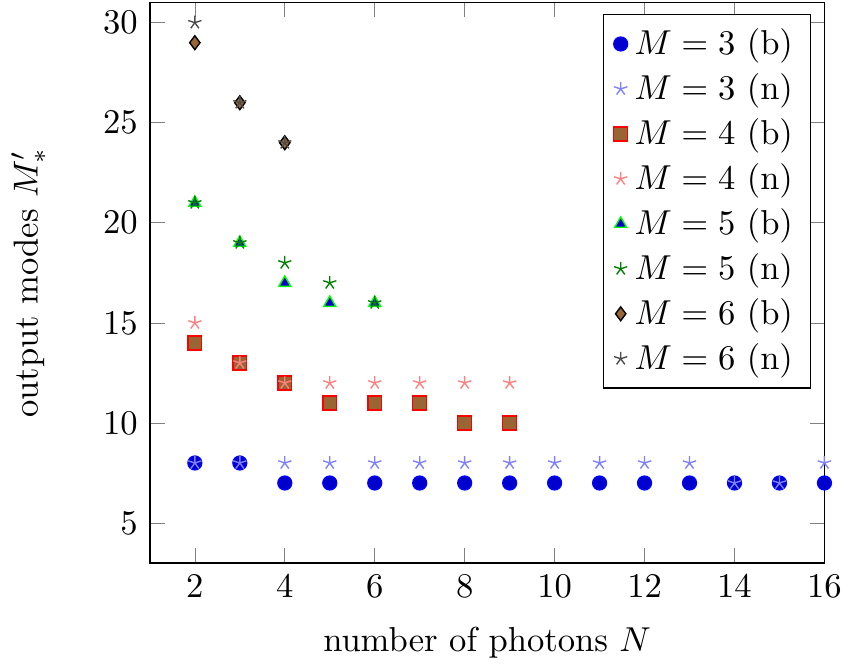}
	\caption{ 
		Number of measurement output modes required for full tomography with a single experimental setup. 
		The lower-bound (b) 
		is estimated from the minimal $M'$ that satisfies Eq.~\eqref{corollary}. 
		The observed numerical value (n) 
		is obtained from the minimal $M'$ such that Eq.~\eqref{eq:pfromrho} 
		is invertible. For $M=2$ we always observe $M'=4$, consistently with Eq.~\eqref{corollary}.
	}
	\label{fig:M1}
\end{figure}

\mysection{Practical implementation}
We have done extensive numerical experiments showing that the bound 
\eqref{count} is achieved by Haar-random configurations
$\{g_\alpha\}_{\alpha=1,\dots,R}$, 
which can be implemented using programmable interferometers
\cite{russell2017direct,burgwal2017using}.
In particular, we find that $\mathcal L$ has full-rank 
$D_{N,M}^2$ only when $R_{N,M}$, or more, configurations are used. 
For $M'>M$, we find that the lower bound \eqref{countp} 
is achievable with $R_{N,M,M'}$, or slightly more, configurations.  
The slightly larger number of configurations or modes required for 
full-tomography when $M'\neq M$ may be due to the simple reconstruction 
algorithm, which does not explicitly take into account independent 
components and normalization.  

The minimum number of measurement modes $M'$ required for a single interferometer is shown in Fig.~\ref{fig:M1}, which shows agreement of numerical results calculated using a single sample from the Haar distribution and the minimal number that satisfies Eq.~\eqref{onelimit}. As predicted by Eq.~\eqref{onelimit2}, 
$M'$ initially decreases as a function of $N$ and then 
becomes constant for $N\approx M$. When $N\approx M$, we find $H_2\approx 1$ 
and hence $M'\gtrsim \alpha M$, thus confirming the scaling relation 
\eqref{onelimit}, and its independence on $N$, although with a larger $\alpha>2$. 
Based on these numerical experiments, we conjecture that with a single 
Haar-random configuration one can perform full-reconstruction with a number of measurement 
modes that increases linearly with $M$. 

In a realistic experiment, the number of detected photons will sometimes fluctuate, 
either because of imperfect photon sources (where $N$-photon states $\rho_N$ are 
generated with probability $\pi_N$), photon losses \cite{garcia2017simulating} 
or imperfect detector efficiency \cite{lee2004towards,achilles2004photon}. 
When there are either imperfect sources or losses, the subset 
of detection events containing exactly the right number of photons 
is sufficient to reconstruct the
state, provided these events occur at an acceptable rate. On the other hand, if losses 
are low and well characterized, one can use all the measured data to reconstruct the entire state
$\rho=\sum_{N=1}^{N_{\rm max}} \pi_N\,\rho_N$ as we show in the Supplementary Material. 

Single-photon detectors (SPDs) that merely distinguish between vacuum and
non-vacuum states are often employed in realistic experiments, instead of true
photon-counting detectors. To achieve sensitivity to photon number, a
nondeterministic number resolving detector (NRD) can be built by multiplexing
SPDs using linear optics and ancillary vacuum states 
\cite{achilles2003fiber,vrehavcek2003multiple,fitch2003photon}. 
We note that this concept is consistent with the scheme shown in 
Fig.~\ref{fig:linearnet2}, and
therefore for sufficiently large $M'$, complete state reconstruction can be
achieved with SPDs. Since an NRD sensitive to $N$ photons requires $N$ SPDs,
Eq.~\eqref{onelimit} implies that $\mathcal O(NM)$ SPDs are required. For $N\ll M$ fewer SPDs
are required, due to the vanishing probability that multiple photons emerge in
the same mode of a random interferometer with $M’>\mathcal O(N^2)$ 
\cite{aaronson2011computational}. More precisely, from Eq.~\eqref{onelimit2} 
we get $M'> \mathcal O(M^2N/\sqrt[N]{N!}) \approx \mathcal O(M^2)$.

\mysection{Conclusion}
We have studied the feasibility and number of measurement configurations required to perform 
quantum tomography of a multi-mode multi-photon Fock state using linear optics and photon counting. 
We have shown that any such state
can be tomographically  reconstructed with a finite number of linear optics 
configurations (Theorem~\ref{thm:feasbility}). To do so, we show that configurations corresponding to any unitary $2N$-design \cite{roy2009unitary} defines an analytical, thought non optimal, reconstruction protocol. Moreover, Theorem~\ref{thm:count} quantifies the minimal number 
of configurations, even when the number of detectors $M'$ 
is larger than $M$.  
For sufficiently many detectors, as specified by Eq.~\eqref{corollary}, 
this leads to tomography with a single measurement configuration.
Our results can be used to test the optimality of tomography 
protocols with a finite number of particles. For instance, 
the two-photon protocol presented in \cite{walser1997measuring}
saturates our bound, and is therefore optimal. Finally, we presented a simple reconstruction algorithm based on 
Haar sampled unitary configurations, and we have observed 
that it is optimal for $M'=M$ and nearly optimality for $M'>M$.

\begin{acknowledgements} \mysection{Acknowledgements} 
The authors thank 
S.~Filippov,
S.~Paesani,
R.~Santagati,
N.~Spagnolo,
B.~Yadin,
for discussions. 
This work is supported by the UK EPSRC grant EP/K034480/1.
MSK thanks the Royal Society,
the KIST Institutional Program (2E26680-18-P025),
and the Samsung GRO grant for their financial support.
\end{acknowledgements}

\appendix

\section{SUPPLEMENTARY MATERIAL}

\section{Bosonic representation of U(M)}\label{a:repr}
Let us consider a set of bosonic creation and annihilation operators 
specified by $a^\dagger_{i\alpha}$ and $a_{j\beta}$. Using second 
quantization notation, we can define the operators
\begin{equation}
	E_{ij} = \sum_\alpha a^\dagger_{i\alpha} a_{j\alpha}~,
	\label{Eope}
\end{equation}
where $i,j=1,\dots,M$, and $\sum_i E_{ii} = N$. 
These operators satisfy the U(M) commutation relations 
\begin{equation}
	[E_{ij},E_{kl}]= \delta_{jk} E_{il}-\delta_{il}E_{kj}~.
\end{equation}
Since bosonic operators are symmetric upon exchange 
of particles, the extra index $\alpha$ will be used to simulate other 
symmetries. For instance, a two-mode anti-symmetric wave function can be 
obtained with $(a_{1H}^\dagger a_{2V}^\dagger - a_{1V}^\dagger a_{2H}^\dagger)\ket 0$,
where $\ket 0$ is the bosonic vacuum and $\alpha=H,V$ is an external index, such 
as the horizontal (H) and vertical (V) photon polarization. 
The operators $H_i := E_{ii}$ are called Cartan operators and form the maximal subset 
of commuting operators in the algebra. Operators $E_{ij}$ are called raising 
operators for $i<j$ and lowering operators for $i>j$. If an eigenstate $\ket\psi$ 
is an eigenstate of all Cartan operators $H_i\ket\psi=\lambda_i\ket\psi$ we
say that the set $(\lambda_1,\dots,\lambda_M)$ is a weight of the state $\ket\psi$. 
Clearly $\sum_i \lambda_i = N$, where $N$ is the number of particles, as $\sum_i H_i= N$. 
Therefore, only $M-1$ Cartan operators $\tilde H_i = H_{i}-H_{i+1}$ are independent. 
A weight $(\lambda_1,\dots,\lambda_M)$  is said to be higher than 
$(\lambda'_1,\dots,\lambda'_M)$  if $\lambda_1>\lambda'_1$, or if $\lambda_1=\lambda_1'$ 
and $\lambda_2>\lambda'_2$ etc. 
Irreducible representations (irreps) of the unitary group are uniquely characterized 
by their highest weight. 
All the other states in a certain irrep can be obtained from the highest weight 
state via multiple applications of the lowering operator \cite{alex2011numerical}. 
A highest weight is a set of 
integers $\lambda_1,\dots,\lambda_M$ that satisfy $\lambda_i\ge \lambda_{i+1}$. 
These numbers can be represented by a Young diagram $Y(\lambda)$, which is an 
array of left adjusted boxes, with $\lambda_1$ boxes on the first row, 
$\lambda_2$ boxes in the second row, etc. 
An alternative labeling of irreps is via Dynkin labels $(b_1b_2\dots)$ where 
$b_j$ is is the number of columns in the Young diagram with $j$ boxes. 

The Gelfand-Tsetlin (GZ) basis \cite{moshinsky1966gelfand} 
is a convenient basis specified by a set of integers 
$m_{ij}$ 
\begin{equation*}
	\begin{pmatrix}
		m_{1,M} & & m_{2,M} & & m_{3,M} & & \dots & & m_{M,M} \\ 
						& m_{1,M-1} && m_{2,M-1} & & \dots && m_{M-1,M-1} \\
						&  &&  & \vdots & &&  \\
						&  &&  m_{1,2} & & m_{2,2} &&  \\
						&  &&  & m_{1,1} & &&  
	\end{pmatrix}.
\end{equation*}
In this basis the $s$-th row $m_{rs}$ gives the irreducible representation of the 
subgroup U(s) in the chain decomposition U(M)$\supset$U(M-1)$\supset\dots\supset$U(1). 
The entries in lower rows satisfy the ``betweenness condition'' $m_{k,i}\ge m_{k,i-1}
\ge m_{k+1,i}$ \cite{alex2011numerical}. 
The first row then specifies the irreps with $\lambda_i = m_{i,M}$ while the other 
rows (that we collectively call $m$) specify the state in that particular irreps. 
These states will be labeled then as $\ket{\lambda, m}$. All Cartan operators $H_i$ 
are diagonal in the GZ basis, with eigenvalues (weights) 
\begin{equation}
	w_i = \sum_{k=1}^i m_{k,i} - \sum_{k=1}^{i-1} m_{k,i-1}~.
	\label{weight}
\end{equation}
Because of the definition Eq.~\eqref{Eope}, the boson number operators are diagonal 
in the GZ basis, namely the GZ basis is like a Fock basis 
but with a convenient labelling. 
Highest weight states 
are such that $m_{ki}=m_{kM}\equiv\lambda_k$, namely where all the elements in the same 
diagonal are equal. These states, that we call $\ket\lambda$, 
have weights $w_i = \lambda_i$ and 
can be written in the second quantized form as 
\cite{moshinsky1966gelfand} 
$\ket{\lambda} \propto B(\lambda) \ket{0}$, 
where $\ket 0$ is the bosonic vacuum, and $B(\lambda)$ is a polynomial 
in the creation operators 
\cite{vilenkin2013representation1}
\begin{equation}
	B(\lambda) = \left(\Delta^1_1\right)^{\lambda_1-\lambda_2} 
		\left(\Delta^{12}_{12}\right)^{\lambda_2-\lambda_3} \cdots
		\left(\Delta^{12\cdots M}_{12\cdots M}\right)^{\lambda_M} ~,
		\label{bosonrep}
\end{equation}
where $\Delta$ is the Slater determinant 
\begin{equation}
	\Delta^{i_1,i_2,\dots}_{j_1,j_2,\dots} = 
	\det \begin{pmatrix}
		a^\dagger_{i_1,j_1} & 
		a^\dagger_{i_1,j_2} & 
		\dots \\
		a^\dagger_{i_2,j_1} & 
		a^\dagger_{i_2,j_2} & 
		\dots \\
		\vdots &\vdots & \ddots
	\end{pmatrix}~.
\end{equation}
The fully symmetric representation with $N$ particles is then specified by 
the highest weight state with 
$\lambda=(N,0,\dots,0)$. 
Other states in the same irrep can be constructed
from the repeated action of lowering operators. For instance, 
$E_{k+1,k} \ket{\{m_{i,j}\}} = \sum_l \alpha_{lk}(\{m\})\ket{\{m_{ij} - \delta_{il}\delta_{jk}\} }$, where the coefficients $\alpha$ are explicitly written in 
\cite{vilenkin2013representation} (Chapter 18.1.2). 

Generally we call a vector $\ket{\lambda,m}$ where $\lambda$ specifies the 
first row in the GZ pattern (and thus defines the irrep), while $m$ 
collects the other rows $m_{k,i}$ for $i<M$. An important vector for 
our discussion is $\ket{\lambda,0}$; using \eqref{weight}, we see that this 
vector is defined by the weight $w_i=0$ for $i<M$ and $w_M=\sum_i \lambda_i =N$. 
Therefore, for the bosonic representation $\lambda=[N]:=(N,0,\dots,0)$ 
this vector corresponds to the ``boson condensate'' state where all the
particles are in the $M$-th mode. Other vectors in $[N]$ are parametrized 
by the GZ pattern 
\begin{equation*}
	\begin{pmatrix}
						 m_{1,M-1} && 0 & & \dots && 0 \\
						  &\ddots&  & \vdots & &&  \\
						  &&  m_{1,2} & & 0 &&  \\
						  &&  & m_{1,1} & &&  
	\end{pmatrix}.
\end{equation*}
where $m_{1,j}$ is related to the occupation number \eqref{weight} via 
$w_i = m_{1,i}-m_{1,i-1}$, so here $w_i$ is the number of particles in mode $i$.

\bigskip

For a given irrep $\lambda$ the Wigner matrices are defined as 
\begin{equation}
	\mathcal D_{m,m'}^\lambda(g) = \bra{\lambda,m}U(g)\ket{\lambda,m'}~.
	\label{wignerd}
\end{equation}
These matrices are orthogonal with respect to the scalar product 
\begin{equation}
	\int dg\, 	\mathcal D_{m,m'}^\lambda(g)  	
	\mathcal D_{\tilde m,\tilde m'}^{\tilde \lambda}(g)^*  = 
	\frac{\delta_{\lambda\tilde\lambda}}{d_\lambda}
	\delta_{m\tilde m}
	\delta_{m'\tilde m'}~,
	\label{ortho}
\end{equation}
where $dg$ is the Haar measure and $d_\lambda$ is the dimension of the representation,
given by \cite{alex2011numerical}
\begin{equation}
	d_{\lambda} = \prod_{1\le k<k'\le M} \left(1+\frac{\lambda_k-\lambda_{k'}}{k'-k}\right)~.
	\label{dim}
\end{equation}

Given a representation $\lambda = (\lambda_1,\dots,\lambda_M)$ the conjugate 
representation can be defined as $\lambda^* = (-\lambda_M,\dots,-\lambda_1)$. 
Indeed, a representation of $g=e^{ih}$ is $U(g)=e^{i \sum_{jk} h_{jk}E_{jk}}$ 
and its conjugate is $U(g)^*=e^{-i \sum_{jk} h^*_{jk}E_{jk}}$, since the operators
\eqref{Eope} are real (in the Fock basis). Therefore, 
the Cartan operators $H_i$ are mapped to $-H_i$ and so are the weights. However,
with these definitions highest weights in the conjugate representation corresponds 
to lowest weights in the original one. The order reversal 
 $\lambda^* = (-\lambda_M,\dots,-\lambda_1)$ assures that highest weights 
 are mapped to highest weights. 
By definition the polynomial $B(\lambda^*)$ associated with the 
conjugate representation can be obtained from \eqref{bosonrep} 
exchanging creation with annihilation operators. 
For each GZ pattern $m$ we can also get the 
corresponding dual pattern $m^*$ by reflecting each row 
$m_{k,i}^* = (-m_{i,i},\dots,-m_{1,i})$.  
Although the GZ pattern now contains negative numbers, this is not a problem 
because two patterns designate the same irrep if $m_{k,M}=m'_{k,M}+c$ -- 
this can be used to bring the GZ basis into the ``normalized'' form 
\cite{alex2011numerical} where $m_{M,M}=0$. However, the polynomials 
\eqref{bosonrep} are defined only when $m_{k,M}\ge 0$.

We consider the tensor product of the symmetric irrep $[N] := (N,0,\dots,0)$ and its 
conjugate $[N]^* := (0,\dots,0,-N)$, which in the normalized form is $(N,\dots,N,0)$. 
As in the addition of angular momenta, this product can be written as a direct 
sum of irreps. In general, see \cite{vilenkin2013representation} Chapter 18.2.6, 
the product of an irrep $\{m_{k,M}\}$ with the fully symmetric one $(N,0,\dots,0)$ 
is a direct sum over the irreps $\{m_{k,M} + p_k\}$ where the non-negative $p_k$ 
satisfy $\sum_k p_k = N$ and 
\begin{align*}
	m_{1,M}+p_1 &\ge 	
	m_{1,M} \ge m_{2,M}+p_2 \ge 	\dots  \ge
	m_{M-2,M} \ge \\&\ge
	m_{M-1,M}+p_{M-1} \ge 	m_{M-1,M} \ge 
	 m_{M,M}+p_{M} \ge 	m_{M,M}~.
\end{align*}
For the product of the symmetric irrep and its conjugate, the above equations force 
$p_2=\dots=p_{M-1}=0$. All the solutions can then be parametrized by an integer 
$0\le \ell\le N$ such that $p_1=\ell$ and $p_M=N-\ell$. The resulting irreps are 
then $(N+\ell,N,\dots,N,N-\ell)$ in the normalized form. These can be written in the
more compact form $\lambda_\ell :=(\ell,0,\dots,0,-\ell)$. Important states for 
our analysis are the ones with zero weight. These are the states such that 
$\sum_{k=1}^i m_{k,i} = 0$, namely the states such that, if you reflect the GZ pattern 
$m$ along the central vertical axis, you obtain $-m$. For $M=2$ and any $\ell$, 
the only state is the one with $m_{1,1}=0$. For $M=3$ all the $\ell+1$ states with 
$m_{1,1}=0$ and $m_{1,2}=-m_{2,2}$ with $0\le m_{1,2}\le \ell$ have zero weight. 
In general the number of zero weight states for fixed $\ell$ is 
\begin{equation}
	d^0_{\ell,M} = \begin{pmatrix}
	\ell+M-2\\\ell 	
	\end{pmatrix}~.
	\label{dimzero}
\end{equation}

\section{Proof of the main theorems}
In Ref.~\cite{christandl2012reliable} (arXiv version) it has been shown that any bosonic 
density matrix (in fact any operator) with $N$ bosons and $M$ modes can be written 
in the integral form (P-representation) 
\begin{equation}
	\rho = \int_{\mathcal U_M} dx \, P_\rho(x)\, \ket{(x)_N}\bra{(x)_N}
	\label{Prep}
\end{equation}
where the integration is over the continuous group
$\mathcal U_M={\rm U}(M)/{\rm U}(M{-}1)\times{\rm U}(1)$, 
\begin{equation}
	\ket{(x)_N} = \ket{x}^{\otimes N} = 
	\frac{\left(\sum_{j=1}^M x_j a_j^\dagger\right)^N}{\sqrt{N!}}\ket 0~,
	\label{condensate}
\end{equation}
is a bosonic condensate, and
\begin{equation}
	P_\rho(x) = \sum_{\ell=0}^N \sum_{m} p_\rho(\ell,m) y_{\ell,m}(x)~,
\end{equation}
where $p_\rho(\ell,m)$ are coefficients, $m$ is a GZ pattern corresponding 
to the irrep $\lambda_\ell$, and $y_{\ell,m}(g) = \mathcal D^{\lambda_\ell}_{m,0}(g)$ 
is written in terms of the Wigner matrices \eqref{wignerd}. 
A generic $g\in{\rm U}(M)$ can be decomposed as $g=xh$ with $x\in\mathcal U_M
$ and $h\in{\rm U}(M-1)\times {\rm U}(1)$. The bosonic condensate \eqref{condensate}
can be written in terms of the state $\ket{\lambda,0}$ for $\lambda=[N]$. Indeed, 
as shown in the previous section, this state is the one where all the particles 
are in the $M$-th mode. Since this state is invariant under U($M-1$) one finds that then 
$\ket{(x)_N} = U(g) \ket{[N],0} \equiv U(x)\ket{[N],0}$. 
Because of this, from the orthogonality of Wigner matrices \eqref{ortho}, one finds that 
\begin{align}
	\int_{\mathcal U_M} \!\!\! dx \, y_{\ell,m}(x) 
y_{\ell',m'}(x)^* &= 	
\int_{{\rm U}(M)} \!\!\!\! dg\, 	\mathcal D_{m,0}^{\lambda_\ell}(g)  	
\mathcal D_{m',0}^{\lambda_{\ell'}}(g)^* = \cr & = \frac1{d_\ell}
\delta_{\ell,\ell'}\delta_{m,m'}~,
\label{orthoy}
\end{align}
where $d_\ell = d_{\lambda_\ell}$. 

Operators diagonal in the Fock basis are also diagonal in the GZ basis. One can therefore 
use the latter for calculations. Off-diagonal elements of the density matrix are then 
\begin{equation}
	\rho_{m,m'} = \bra{[N],m}\rho\ket{[N],m'}~.
\end{equation}
Using the P-representation \eqref{Prep}, one has 
$\bra{[N],m}(x)_N\rangle = \mathcal D^{[N]}_{m,0}(x)$, and similarly
$\langle(x)_N\ket{[N],m} = (\bra{[N],m}(x)_N\rangle)^* = \mathcal D^{[N]}_{m,0}(x)^*=
\mathcal D^{[N]^*}_{m^*,0}(x)$,
so 
\begin{align}
	\rho_{m,m'} &= 
	 \int dx \, P_\rho(x)\, 
\mathcal D^{[N]}_{m,0}(x)
\mathcal D^{[N]^*}_{m'{}^*,0}(x)
\\&=
\int dx \, P_\rho(x)\, \left[
\mathcal D^{[N]^*}_{m^*,0}(x)
\mathcal D^{[N]}_{m',0}(x)\right]^*~.
\end{align}
An explicit form for the polynomial $\mathcal D^{[N]}_{m,0}(x)$
for $x\in \mathcal U_M$ is written 
in \cite{vilenkin2013representation1} (Chapter 5.2.5) 
\begin{equation}
	\mathcal D^{[N]}_{m,0}(x) =  \frac1{\sqrt{d_{[N]}} }
	\prod_{j=1}^M \frac{x_j^{w_j}}{\sqrt{w_j!}}
	\label{dexample}
\end{equation}
where $w_j=m_{1,j}-m_{1,j-1}$. 
As we have shown in the previous section the tensor product of $[N]$ and $[N]^*$ 
is a sum of irreps $\lambda_\ell$. From the expansion of 
Wigner functions (see \cite{vilenkin2013representation} Chapter 
18.2.1)
\begin{align}
	\mathcal Y_{m',m^*}(x) &= 
	\mathcal D^{[N]^*}_{m^*,0}(x)
	\mathcal D^{[N]}_{m',0}(x) \cr &= \sum_{\ell=0}^N \sum_{\tilde m,m_0}  
	\Gamma^N_\ell(m',m^*|\tilde m,m_0)
	\mathcal D^{\lambda_\ell}_{\tilde m,m_0}(x)
	 \label{DtoY}
\end{align}
where 
\begin{align}
	\Gamma^N_\ell(m',m^*|\tilde m,m_0) = \sum_r &
	\bra{[N]^*,m^*;[N],m'}\lambda_\ell,r,\tilde m\rangle\times \cr & \times
\langle \lambda_\ell,r,m_0 \ket{[N]^*,0;[N],0}
\end{align}
and $\bra{\lambda,m;\lambda',m'}\lambda'',r,m''\rangle$ is the Clebsch-Gordan coefficient 
with multiplicity $r$ \cite{vilenkin2013representation,alex2011numerical}. 
The Clebsch-Gordan coefficients are real, and 
different from zero only if the weights \eqref{weight}
coincide \cite{alex2011numerical}, namely if $w_i(m')+w_i(m) = w_i(m'')$ for any $i$. 
When $m'=m^*$ the only possibility is when $m$ is a zero weight state. 
Therefore, from the orthogonality \eqref{orthoy} we find 
\begin{align}
	\rho_{m,m'} &= \int_{{\rm U}(M)} dg \, P_\rho(g)\, \mathcal Y^*_{m',m^*}(g)
						= \cr & = 
	\sum_{\ell=0}^N \sum_{\tilde m}  \frac{p_\rho(\ell,\tilde m)}{d_\ell}
	\Gamma^N_\ell(m',m^*|\tilde m,0) ~,
	\label{offdiag}
\\
	\rho_{m,m} &= \sum_{\ell=0}^N \sum_{\tilde m\in Z_\ell}  
	\frac{p_\rho(\ell,\tilde m)}{d_\ell}
	\Gamma^N_\ell(m,m^*|\tilde m,0) ~,
	\label{diagz}
\end{align}
where we used \eqref{DtoY} and the orthogonality \eqref{ortho}. 
In \eqref{offdiag}
where the range of $\tilde m$ is bounded for any $\ell$ by the 
``betweenness'' condition, and $Z_\ell$ is the set of zero weight states 
in $\ell$. From the dimensionality \eqref{dimzero}, and since 
$\sum_{\ell=0}^N d^0_{\ell,M} = D_{N,M}$ we see that the diagonal elements 
of the density matrix are in one-to-one correspondence with the numbers 
$p_\rho(\ell,m)$ for $m\in Z_\ell$.

We now calculate the diagonal elements after the application of a rotation $g'\in{\rm U}(M)$,
\begin{align}
	 \label{diag}
	p_{n}(g') &=  \bra{[N],n}U(g)\rho U(g)^\dagger \ket{[N],n}
				= \\\nonumber & =
				\int_{{\rm U}(M)} dg \, P_\rho(g)\, \mathcal Y^*_{n,n^*}(g'{}^\dagger g)~.
\end{align}
Since 
\begin{align}
	\mathcal{D}^{\lambda_\ell}_{\tilde m,m_0}(g'{}^\dagger g)^*
	 &=  \sum_{\lambda,m} 
	\bra{\lambda_\ell,\tilde m }U(g')^\dagger\ket{\lambda,m}^*
	\bra{\lambda,m}U(g)\ket{\lambda_\ell,m_0}^*
	 \cr &= \sum_m
	 \mathcal{D}^{\lambda_\ell}_{m,\tilde m}(g') 
	 \mathcal{D}^{\lambda_\ell}_{m,m_0}(g)^*
\end{align}
where in the first equation we introduce a resolution of the identity. 
%
Inserting the above equation in \eqref{diag} and using the orthogonality 
\eqref{ortho} we find 
\begin{align}
	p_{n}(g) &  
	= \sum_{\ell=0}^N \sum_m \sum_{m_0\in Z_\ell}	\Gamma^N_\ell(n,n^*|m_0,0) 
 \frac{p_\rho(\ell,m)}{d_\ell} 
 \mathcal D^{\lambda_\ell}_{m,m_0}(g)~.
	 \label{diagnoint}
\end{align}
The Clebsch-Gordon coefficients satisfy the orthogonality relations 
\cite{vilenkin2013representation}
\begin{align*}
	\sum_{mm'}
	\langle\mu',r',n'\ket{\lambda,m;\lambda',m'}
	\bra{\lambda,m;\lambda',m'}\mu,r,n \rangle
	&= \delta_{\mu,\mu'}\delta_{nn'}\delta_{rr'}~,
	\\
	\sum_{\mu r p}
	\bra{\lambda,m;\lambda',m'}\mu,r,p \rangle
	\langle\mu,r,p\ket{\lambda,n;\lambda',n'}&= \delta_{m,n}\delta_{m'n'}~.
\end{align*}
From these we find the orthogonality relation 
\begin{align}\nonumber
	\sum_{m,m'} \Gamma^N_\ell(m,m'&|\tilde m,m_0) 
	\Gamma^N_{\ell'}(m,m'|\tilde m',m_0') = \delta_{\ell,\ell'} \times \\
																					&\times 
	\delta_{\tilde m,\tilde m'}\Gamma^N_\ell(0,0|m_0,m_0')~.
	\label{e.cg1}
\end{align}
and 
\begin{equation}
	\sum_{\ell, \tilde m} \Gamma^N_\ell(m,m'|\tilde m,\tilde m) = \delta_{m,0}
	\delta_{m',0}~.
\end{equation}
Moreover, when $\tilde m\in Z_\ell$ the selection rule shows that for each 
$m$ in $[N]$ there is only a single $m'$ in $[N]^*$ such that 
$w_j(m)+w_j(m')=0$; this state is $m'=m^*$. Therefore, we can remove 
one element from the sum \eqref{e.cg1} and write 
\begin{align*}
	\sum_{n} \Gamma^N_\ell(n,n^*&|\tilde m_0,m_0) 
	\Gamma^N_{\ell'}(n,n^*|\tilde m_0',m_0') = \delta_{\ell,\ell'} \times \\
																					&\times 
	\delta_{\tilde m_0,\tilde m_0'}\Gamma^N_\ell(0,0|m_0,m_0')~,
\end{align*}
for $\tilde m_0 \in Z_\ell, \tilde m_0'\in Z_{\ell'}$. From the above relations 
\begin{align*}
	\sum_n \Gamma^N_\ell(n,n^*|m_0,0)  p_{n}(g)  
	= \sum_m	 \frac{P_\rho(\ell,m)}{d_\ell} 
	\Gamma^N_\ell(0,0|0,0)  
 \mathcal D^{\lambda_\ell}_{m,m_0}(g)~.
\end{align*}
Since $\Gamma^N_\ell(0,0|0,0)\neq0$ 
for any $0\le\ell\le N$ (\cite{christandl2012reliable} Corollary 3),
we define then 
\begin{equation}
	p^\ell_{m_0,g} = \sum_n 	
	\frac{\Gamma^N_\ell(n,n^*|m_0,0)^*}{ \Gamma^N_\ell(0,0|0,0) } p_n(g) = 
 \sum_m 	
 \frac{P_\rho(\ell,m)}{d_\ell} 
 \mathcal D^{\lambda_\ell}_{m,m_0}(g)~.
	 \label{diagnointl}
\end{equation}
The above result shows that if we can find a discrete set of unitaries $g_\alpha$ such that 
$ \mathcal D^{\lambda_\ell}_{m,m_0}(g)$ 
is invertible, then we can obtain $P_{\rho}(\ell,m)$ and 
hence $\rho_{m,m'}$. Indeed, if there is a discrete set of unitaries 
$\{g_\alpha\}$ and a matrix $X$ such that
\begin{equation}
	\sum_{m_0,\alpha} X^\ell_{m',m_0g_\alpha}  \mathcal D^{\lambda_\ell}_{m,m_0}(g_\alpha)
	=\delta_{m,m'}~,
	\label{Xsol}
\end{equation}
then from Eq.~\eqref{offdiag}
\begin{align}
	\rho_{m,m'} &= 
	\sum_{\ell=0}^N \sum_{\tilde m,m_0,\alpha}  
	\Gamma^N_\ell(m',m^*|\tilde m,0) X^\ell_{\tilde m,m_0g_\alpha} 
	p^\ell_{m_0,g_\alpha} 
	\label{tomoX}
	~.
\end{align}

\subsection{Proof of Theorem \ref{thm:feasbility}: Unitary design}\label{a:design}
An analytic solution to Eq.~\eqref{Xsol} is obtained from the 
theory of unitary designs \cite{roy2009unitary}. A set of unitaries 
${g_\alpha}_{\alpha=1,\dots,R}$ is called a {\it weighted unitary 
$t$-design} if 
\begin{equation}
	\sum_\alpha w(g_\alpha) g_\alpha^{\otimes t} \otimes g_\alpha^{*\otimes t} 
	= \int_{U(M)} dg \, g^{\otimes t} \otimes g^{*\otimes t}~,
\end{equation}
where $dg$ is the Haar measure, and $w$ is a weight. 
In other terms, a weighted unitary $t$-design is a collection of unitaries such 
that the weighted average of any polynomial function $f_t(g)$, 
with maximal degree $t$ in both $g_{ij}$ and $g_{kl}^*$, is equal to the 
average over the entire continuous group $\int dg \, f_t(g)$. 
From the expansion \eqref{DtoY} and from \eqref{dexample}, we see that 
$\mathcal D_{m,m_0}^{\lambda_\ell}$, and in particular 
$\mathcal D_{m,m_0}^{\lambda_N}$, are at most polynomial functions of 
degree $N$ in $g$ and $g^*$. Therefore, if ${g_\alpha}_{\alpha=1,\dots,R}$ 
is a weighted $2N$-design, then a solution of \eqref{Xsol} is obtained 
by setting $X^\ell_{m,m0,g} = d_\ell w(g) \mathcal D_{m,m_0}^{\lambda_\ell *}$. Indeed, 
from the definition of unitary design Eq.~\eqref{Xsol} becomes 
\begin{align*}
	\sum_{m_0,\alpha} X^\ell_{m',m_0g_\alpha}  \mathcal D^{\lambda_\ell}_{m,m_0}(g)&=
	d_\ell \sum_{m_0,\alpha} w(g_\alpha) \mathcal D^{\lambda_\ell}_{m',m_0}(g_\alpha)^*
	\mathcal D^{\lambda_\ell}_{m,m_0}(g_\alpha)
	\cr & = d_\ell
	\int dg\, \mathcal D^{\lambda_\ell}_{m',m_0}(g)^*
	\mathcal D^{\lambda_\ell}_{m,m_0}(g)
	=\delta_{m,m'}~,
\end{align*}
where in the last equation we used the orthogonality of Wigner matrices. 
Therefore, for any weighted unitary $2N$-design, one can find an analytic 
solution to the tomographic reconstruction of the state
\begin{align*}
	\rho_{m,m'} &= 
	\sum_{\ell=0}^N \sum_{\tilde m,m_0,\alpha}  
	\Gamma^N_\ell(m',m^*|\tilde m,0) d_\ell
w(g_\alpha) \mathcal D^{\lambda_\ell}_{\tilde m,m_0}(g_\alpha)^*
	p^\ell_{m_0,g_\alpha} 
	~.
\end{align*}
Note that the requirement that the matrices $g_\alpha$ form a 
unitary $2N$-design is much stronger than \eqref{Xsol}.
Indeed, with a unitary design, the average over {\it any} polynomial function
is equal to the group integral, while for inverting \eqref{Xsol}  it is 
required that the discrete average is equal to the group integral only
for specific set of functions (product of Wigner matrices). 
Upper and lower bounds on the size of a $2N$-design is given in \cite{roy2009unitary}, where
they found that 
\begin{equation}
	B(M,N) \le R\le B(M,2N)~,
	\label{royscott}
\end{equation}
where 
\begin{equation}
	B(M,N) = \sum_{\lambda:|\lambda_+|\le N} d_{\lambda,M}^2~.
\end{equation}
where $|\lambda_+|$ is the sum of the positive elements in $\lambda$
and $d_{\lambda,M}$, with explicit dependence on $M$, is written 
in \eqref{dim}. Moreover, \cite{roy2009unitary}
\begin{equation}
	B(M,N) \le \binom{M^2+N-1}{N}^2 \equiv D_{N,M^2}^2.
\end{equation}

\subsection{Proof of Theorem~\ref{thm:count}}

Let $R$ be the number of unitaries ${g_\alpha}_{\alpha=1,\dots,R}$. 
The matrix $Y^\ell_{m,m0\alpha} = 
\mathcal D^{\lambda_\ell}_{m,m_0}(g_\alpha)$ has dimension 
$d_\ell\times(R d_{\ell,M}^0)$.  As in the main text, we focus 
on the higher dimensional irrep, namely $\ell=N$, 
and we want to count the number $R$ of matrices $g_\alpha$ required
to make the matrix $Y^N$ invertible.  Note that the Wigner $D$ functions
$\mathcal D^{\lambda_\ell}_{m,m_0}(g_\alpha)$ 
play the same role of the irreducible symmetric tensor of rank $r=\ell$ 
discussed in the main text. For $r=N$, thanks to 
Eq.~\eqref{dimzero}, we find  $d_{N,M}^0 = D_{N,M-1}$. 
On the other hand, explicitly solving Eq.~\eqref{dim}, we find 
$d_N = R_{N,M} D_{N,M-1}$. Therefore, in order to make $Y$ invertible for 
$\ell=N$, one finds $R\ge R_{N,M}$. This completes the proof of 
theorem ~\ref{thm:count}, as the inverse in \eqref{Xsol} exists. 
The lower bound can be achieved if the 
same unitaries $g_\alpha$ enable also the inversion $Y^\ell$ for 
$\ell<N$. The reconstruction algorithm is then Eq.~\eqref{tomoX}. 


Finally, 
we  consider the case where the number of output modes $M'$ is different 
from the number of inputs. Technically, all equations are still valid, because 
we can first expand the input space into $M'>M$ modes, and select only 
inputs coming from $M$ mode subspace. One can still write then Eq.~\eqref{Xsol} 
and \eqref{tomoX}. As for Eq.~\eqref{Xsol}, for the worst case scenario $\ell=N$,
the number of possible $m_0$ is now 
$d^0_{N,M'} = D_{N,M'-1}$, while the number of $m$ is $d_N = R_{N,M}D_{N,M-1}$. 
From this we find an alternative proof our corollary Eq.~\eqref{corollary}.

\subsection{Analytical solutions for M=2} \label{s:m2}
SU(2) is formed by the matrices 
\begin{equation}
	U(g) = e^{i\phi S_z} e^{i\theta S_y} e^{i\eta S_z}~,
\end{equation}
where $S_\alpha$ are the spin matrices in an arbitrary representation. The
states $\ket{\lambda_l,m}$ are parametrized by a single integer $-\ell\le m\le \ell$. 
Moreover, there is only one zero-weight state $m_0=0$. 
To map these indices to standard spin notation we parametrize these numbers with semi-integers 
$S=\ell/2$ and $M=m/2$ (in this section $M$ is not the number of modes, which is always two), 
therefore
\begin{equation}
	\mathcal D^{\lambda_\ell}_{m,m_0}(g) = y_{\ell,m}(g) = e^{i \phi M} d_{M,0}^S(\theta)~,
\end{equation}
where $d_{M,M'}^S(\theta) = \bra{S,M}e^{i\theta S_y}\ket{S,M'}$ is the Wigner function,
and the dependence of $\eta$ disappears. 
If we choose $X^S_{M}(\theta,\phi) = \gamma e^{-i \phi M} [d_{M,0}^S(\theta)]^{-1}$,
where $\gamma$ is a normalization, and hence 
we need to select $\theta$ such that $d_{M,0}^S(\theta) \neq 0$ then the
condition \eqref{Xsol} can be written as 
\begin{equation}
	\gamma \sum_\alpha e^{i \phi_\alpha (M-M')} 
	\frac{d_{M,0}^S(\theta_\alpha)}{d_{M',0}^{S}(\theta_\alpha)} = 
	\delta_{M,M'}~,
\end{equation}
where $-2S\le M-M'\le 2S$, namely $-\ell\le M-M'\le \ell$. 
In the worst case scenario, $-N\le M-M'\le N$. 
Therefore, using standard properties of discrete Fourier Transform, 
a solution of the above equation, valid for any $M$ and $M'$, is 
\begin{align}
	\theta_\alpha&=\theta~, &  \phi_\alpha &= \frac{2 \pi \alpha}{2N+1}~,
							 & \gamma = \frac1{2N+1}~,
\end{align}
for $\alpha=0,\dots,2N$, where $\theta$ is such that  $d_{M,0}^S(\theta) \neq 0$
for any $M$. This solution is therefore equivalent to the 
Newton and Young algorithm for SU(2) tomography \cite{newton1968measurability}. 
As the number of setups is $R=2N+1$, this 
protocol saturates the bound \eqref{count}. 

\section{Alternative proof of the minimal number of configurations }

We first present an alternative proof of Eq.~\eqref{count}.  
We note that photodetection 
is equivalent to the measurement of all the expectation values of polynomials of 
number operators $T^k_k = a^\dagger_ka_k$. Not all of these expectation values 
are independent. 
For instance, if one measures $\mean{T^k_k}$ for $k=1,\dots,M-1$, then one gets 
$\mean{T^M_M}=N-\sum_{k=1}^{M-1}\mean{T_k^k}$ without further measurements. 
Thanks to Wick's theorem,
all independent polynomials in the number operators can be written via the rank $r$ tensors 
$T^{\alpha_1,\dots,\alpha_r}_{\alpha_1,\dots,\alpha_r} = 
a^\dagger_{\alpha_1}\cdots a^\dagger_{\alpha_r} a_{\alpha_1}\cdots a_{\alpha_r}$,
where we fix the convention that Greek indices $\alpha_j$ range from 1 to $M-1$ while 
Latin indices $k_j$ range from 1 to $M$. 
Tensors with some indices $M$ can be written in terms of other tensors 
with indices from 1 to $M-1$. For fixed rank $r$, the number of independent expectation 
values $\mean{T^{\alpha_1,\dots,\alpha_r}_{\alpha_1,\dots,\alpha_r}}$ is $D_{r,M-1}$. 
Moreover, such expectation values are zero for $r>N$. The expectation values 
$\mean{T^{\alpha_1,\dots,\alpha_r}_{\alpha_1,\dots,\alpha_r}}$ for $r=1,\dots,N$ 
uniquely specify the outcome of photodetection. Indeed, 
$\sum_{r=1}^N D_{r,M-1} = D_{N,M}-1$ so all the information about photodetection is 
contained in the expectation values of the independent tensors 
$T^{\alpha_1,\dots,\alpha_r}_{\alpha_1,\dots,\alpha_r}$. 

Consider now a linear optical transformation, 
namely a collection of beam splitters and phase shifters,
expressed by an 
irreducible representation $U(g)=e^{i \sum_{kl} H_{kl} a_k^\dagger a_l}$ of a 
SU(M), where $g=e^{i H}$ is a $M\times M$ unitary matrix. 
Applying first a linear optical network and then performing 
photodetection is equivalent to the measurement of the diagonal elements of 
$U(g)\rho U(g)^\dagger$ or, equivalently in the Heisenberg picture, 
of the independent tensors 
$U(g)^\dagger T^{\alpha_1,\dots,\alpha_r}_{\alpha_1,\dots,\alpha_r} U(g)$. 
With many choices of $g$ one then aims at measuring all the rank $r$ tensors 
\begin{equation}
T^{k_1,\dots,k_r}_{\ell_1,\dots,\ell_r} = 
a^\dagger_{k_1}\cdots a^\dagger_{k_r} a_{\ell_1}\cdots a_{\ell_r}~,
\label{T}
\end{equation}
for $1\le r\le N$. 
Note that the above tensor has Latin indices, each ranging from $1$ to $M$. In fact 
$U(g)^\dagger T^{\alpha_1,\dots,\alpha_r}_{\alpha_1,\dots,\alpha_r} U(g)$ may have 
some indices equal to $M$. The above tensor is totally symmetric in its upper and lower 
indices, but not in mixtures of them. The numbers
$\mean{T^{k_1,\dots,k_r}_{\ell_1,\dots,\ell_r}}$ provide $D_{r,M}^2$ expectation values, 
but not all of them are independent, since 
$\mean{T^{M,k_2,\dots,k_r}_{M,\ell_2,\dots,\ell_r}}$ can be written in terms of other 
expectation values. These dependent operators are $D_{r-1,M}^2$. Therefore, 
the number of independent rank $r$ expectation values is
$D_{r,M}^2-D_{r-1,M}^2$. The independent expectation values are the numbers
$\mean{T^{k_1,\dots,k_r}_{\ell_1,\dots,\ell_r}}$ where there is 
never both an upper and lower index $M$. These numbers 
completely and uniquely specify the state $\rho$, and indeed 
$\sum_{r=1}^N D_{r,M}^2-D_{r-1,M}^2 = D_{N,M}^2-1$, which is the number of independent 
components in the density matrix $\rho$. Rank $r$ tensors transform as 
$U(g)^\dagger T^{k_1,\dots,k_r}_{\ell_1,\dots,\ell_r}U(g) = [g^{\dagger \otimes r} T 
g^{\otimes r }]^{k_1,\dots,k_r}_{\ell_1,\dots,\ell_r}$, so each photodetection 
after the linear transformation $g$ allows us to measure the independent components 
$ [g^{\dagger \otimes r} T 
g^{\otimes r }]^{\alpha_1,\dots,\alpha_r}_{\alpha_1,\dots,\alpha_r}$. What is 
the minimal number of $g$ to reconstruct all off-diagonal elements 
$\mean{T^{k_1,\dots,k_r}_{\ell_1,\dots,\ell_r}}$? Fox fixed $r$ all the tensor components 
are independent (provided that index $M$ is not both on upper and lower indices), so the most 
difficult tensor elements to reconstruct are those for $r=N$. The number of independent 
components in this tensor is $D_{N,M}^2-D_{N-1,M}^2$, but each photodetection with 
a fixed choice of $g$ allows us to 
get $D_{N,M-1}$ independent values. Therefore, the minimal number of $g$ is 
$[D_{N,M}^2-D_{N-1,M}^2]/D_{N,M-1} \equiv R_{N,M}$. This concludes the proof of  
Eq.~\eqref{count}. Similarly, Eq.~\eqref{countp} appears for different number of modes 
as $R_{N,M,M'}=[(D_{N,M}^2-D_{N-1,M}^2)/D_{N,M'-1}]\equiv [R_{N,M} D_{N,M-1}/D_{N,M'-1}]$.

Note that this construction does not show that this lower bound 
is achievable, because the values of $g$ used to construct the rank $N$ expectation values 
may not enable the reconstruction of the other expectation values with $r<N$. 
Nonetheless, an explicit protocol which achieves our lower bound is known 
\cite{walser1997measuring}  for $M=2$.

\subsection{Analytical reconstruction algorithm for $M=2$}
It is instructive to rephrase the analytic 
protocol discussed in Section~\ref{s:m2} for $M=2$, 
based on the spin tomography protocols \cite{newton1968measurability}, 
to see 
why the settings developed for the reconstruction of rank $N$ tensors are normally 
enough for the reconstruction of lower-rank tensors.  
According to \cite{newton1968measurability}, SU(2) tomography can be achieved 
with $R_{N,2}=2N+1$ unitary matrices $g$, obtained by first applying 
different rotations $\phi_j$ along the $z$ axis, with 
$\phi_j = 2\pi j/(2N+1)$ and $j=1,\dots,2N+1$, and then a fixed
rotation with angle $\theta$ along the $y$ axis  
(see also Appendix \ref{s:m2}). 
In the linear optics setup, this corresponds to the application of 
different phase shifts $\phi_j$ on a single mode, followed by a beam splitter
with transmissivity related to $\theta$. We set then 
$U(g)=e^{\theta (a_1^\dagger a_2-a_2^\dagger a_1)} e^{i\phi a_2^\dagger a_2} $
and we note that for $M=2$ the Greek indices can only take the single value $\alpha_j=1$.
Therefore, $ p(\phi,\theta)=\mean{[g^{\dagger \otimes r} T 
g^{\otimes r }]^{\alpha_1,\dots,\alpha_r}_{\alpha_1,\dots,\alpha_r}}
= \sum_{\{k_j\},\{\ell_j\}}\mean{T^{k_1,\dots,k_r}_{\ell_1,\dots,\ell_r}}
\prod_{j=1}^r (g_{k_j,1}^*g_{\ell_j,1} )
$, where the left-hand side contains the measured values and the right hand side 
contains the off-diagonal elements that we want to reconstruct. 
Calling $f_{k\ell}(\theta)$ the $\theta$ dependent part, we can write 
$ p(\phi,\theta)=\sum_{\{k_j\},\{\ell_j\}} f_{k\ell}(\theta)
\mean{T^{k_1,\dots,k_r}_{\ell_1,\dots,\ell_r}} e^{i\phi (n_{\{\ell{=}2\}}-n_{\{k{=}2\}})}$,
where $n_{\{\ell{=}2\}}$ is the number of indices $\ell_j=2$ and similarly for 
$n_{\{k{=}2\}}$. Since independent tensors cannot have indices equal to 2 
in both upper and lower indices, the number $I=n_{\{\ell{=}2\}}-n_{\{k{=}2\}}=-r,\dots,r$ 
completely specify the indices of the independent tensors. To reconstruct 
$\mean{T^{k_1,\dots,k_r}_{\ell_1,\dots,\ell_r}}$ one then simply has to choose 
different phases $\phi_j$ such that the matrix $e^{i\phi_j I}$ with 
$I=-r,\dots,r$ can be inverted for any $r$. Clearly one can consider the worst 
case $r=N$ and chose the Fourier transform where  $\phi_j = 2\pi j/(2N+1)$, so 
that $\mean{T(I)} \propto \sum_j e^{-i\phi_j I} p(\phi_j,\theta)/f_I(\theta)$,
where $I$ groups the indices $k$ and $\ell$.  We remark that this
choice is obtained by trying to invert the most complicated case $r=N$, but since
$\sum_j e^{i\phi_j (I-J)}\propto \delta_{I,J}$ for any $-N\le I,J\le N$, with the 
same rotations one automatically obtains the lower rank tensors, where the only difference is 
that $I,J$ are constrained to smaller ranges.  
We conjecture that a similar construction applies also for higher values of $M$, and that
the bound $R_{N,M}$ can always be achieved. This is indeed what we have observed in 
numerical simulations for different values of $M$ and $N$ and $R_{N,M}$ 
Haar-random choices of $g$.

\section{Imperfections} 
Here we study in more detail how to deal with possible errors in photon sources  and 
detectors.  Eventual photon losses in the interferometer can be included into the 
detector efficiency. 
Indeed, consider a model of an imperfect measurement as an extended linear optical interferometer
wherein some ancillary modes are unmeasured \cite{garcia2017simulating}. 
For uniform efficiency per mode $\eta$, the input-output relationship 
\cite{garcia2017simulating} is described by $\eta U$ where $U$ is a unitary 
$M' \times M'$ matrix (as in Fig.~1 of the main text). The resulting
probability, given by Eq.~2 of the main text, is then 
$p_{\nu',g,\eta} = \eta^N p_{\nu',g}$ where $\eta^N$ is the probability of 
loosing no photons, and $p_{\nu',g}$ is the conditional output probability,
given that no photons have been lost. Because of this, photon losses can be 
modeled via auxiliary beam splitters that bring  some photons to an unmeasured environment. 
An imperfect detector can also be modeled as a perfect detector with an extra beam splitter 
in front. Therefore, assuming that losses are uniform and independent on the configuration 
of the interferometer, we can include them into the detector efficiency. 

We first consider the simplified case where either the photon sources or the detectors 
are imperfect, where a simple post-selection can be applied to reconstruct the correct state. 
The case where both imperfect sources and detectors are present is treated 
then in Sec. \ref{s:im}.

\subsection{Imperfect sources, perfect detectors} \label{s:imperfectsources}
Suppose that the source emits $N$ photons with probability $\pi_N$. 
The resulting state is then 
\begin{equation}
	\rho = \sum_{N=0}^{N_{\rm max}} \pi_N  \rho_N~,
	\label{e:rhomultiN}
\end{equation}
where each $\rho_{N}$ is a possibly mixed state with $N$ photons in $M$ modes. 
This probabilistic description is appropriate for common source imperfections, 
such as emission of a single emitter into an undesired spatial mode and 
inefficient heralding from a two-mode state generated by spontaneous parametric wave-mixing.
 Assuming perfect detectors, our protocol can be trivially extended to measure not only $\rho_N$ 
for the expected number of photon $N$, but also the entire state $\rho$ with the 
resulting probabilities $\pi_N$.  Indeed, we can expand the photon probability after detection 
as 
\begin{align}
	p_n(g) &= \bra{n}U(g)\rho U(g)^\dagger\ket n \cr &= \sum_N \pi_N \, 
	\bra{[N],n}U(g)\rho_N U(g)^\dagger\ket{[N],n}
	\cr &= \sum_n \pi_N p^{N}_n(g) ~,
	\label{e:probnN}
\end{align}
where $\ket{n}=\ket{n_1,\dots,n_m}$ is a generic multimode Fock state and 
$\ket{[N],n}$ is a Fock state such that $\sum_j n_j=N$. By post-selecting 
the measurement outcomes such that there are exactly $N$ photons one can reconstruct 
the conditional probability 
\begin{align}
	p_n^N(g) = \bra{[N],n}U(g)\rho_N U(g)^\dagger\ket{[N],n}~.
\end{align}
These probabilities can then be used to reconstruct the state following the procedure 
outlined in the main text. In other terms, post-selection allows us to treat the 
photon state in Eq.~\eqref{e:rhomultiN} as having exactly $N$ particles. The number 
of measurement settings is then $R_{N,M}$, or $R_{N,M,M'}$ when the number of output 
modes is different. 
Nonetheless, one can go much further: 
by post-selecting over different photon numbers one can then reconstruct all the 
states $\rho_N$ for $N=1,\dots,N_{\rm max}$. From the same outcomes one can then 
estimate $\pi_N$, as the posterior probability of detecting $N$ photons. 

We have observed in all numerical experiments 
that the same settings $g$ used to reconstruct $\rho_N$ 
can also be used to reconstruct the other states $\rho_{N'}$ with $N'\neq N$. 
Assuming that this numerical observation holds true in general, 
the required number of settings for complete reconstruction  is then 
\begin{equation}
	\max_{N\in \{1,\dots,N_{\rm max}\}} R_{N,M,M'}~.
\end{equation}
The above number also provides the lower bound on the number of settings required 
for state tomography when the number of photons is uncertain. 
Remarkably, this number can be finite even when $N_{\rm max}$ is unbounded. 
This is due to the fact that, as we have discussed in the main text and shown in Fig.~(2), 
$R_{N,M,M'}$ can decrease as a function of $N$, for certain choices of $M'$.

\subsection{Perfect sources, imperfect detectors}

Consider a single detector with efficiency $\eta$. The probability to detect $k$ photons 
when there are $n$ incident photons is given by 
\begin{equation}
	P_\eta(k|n) = \binom{n}{k}\;\eta^k(1-\eta)^{n-k}~.
\end{equation}
A simple way to explain the above formula is that a photodetector with its detection
efficiency $\eta$ can be interpreted as a perfect detector with a beam splitter
in front. Here the beam splitter’s transitivity  is determined by the
detection efficiency. In case of multiple detectors $n=\{n_1,\dots,n_M\}$, 
$k=\{k_1,\dots,k_M\}$  and 
\begin{equation}
	P_\eta(k|n) = \prod_{j=1}^M P_{\eta_j}(k_j|n_j)~,
\end{equation}
where $\eta_j$ is the efficiency of the detector on mode $j$. 

The simplest way to deal with imperfect detectors is again post-selection: assuming 
that the input state has exactly $N$ photons, one can ignore all  measurement outcomes 
where the total number of detected photons is different from $N$. This approach is viable 
as long as the data rate is sufficient. 

Nonetheless, it has been shown in \cite{lee2004towards,achilles2004photon} 
that neglecting outcomes (namely post-selection) is not necessary, as long 
as the detection efficiency is well characterized. Indeed, suppose that 
the true photon distribution is $p^{N}_n(g)$, where $\sum_j n_j=N$, then 
the detected distribution is 
\begin{equation}
	p^d(k,g) = \sum_n P_\eta(k|n) p^N_n(g)~.
	\label{e:detect}
\end{equation}
Different methods have been proposed \cite{lee2004towards,achilles2004photon} to 
reconstruct $p^N_n(g)$ given the detected probability $p^d(k,g)$, e.g. based on 
inverting the above matrix equation, or with maximum likelihood estimators or finally 
using Bayes' theorem. 

\subsection{Imperfect sources, imperfect detectors}\label{s:im}
When there are both imperfect sources and imperfect detectors, then post-selection is 
not anymore a viable solution. Suppose indeed that the initial state is the one of 
Eq.~\eqref{e:rhomultiN}, where the expected number of photons is $N$, while larger or 
smaller values are due to imperfect sources. It could be that the source generates 
a higher photon-number, but then one photon is lost so the number of detected photons is still
$N$. Post-selection is clearly not able to remove this source of errors, so one has to rely on
the inversion of Eq.\eqref{e:detect}, which is expected to be highly accurate when 
the detectors' efficiency is sufficiently high. 

More precisely, we can combine \eqref{e:probnN} with \eqref{e:detect} to write 
\begin{align}
	\label{firstline}
	p^d(k,g) &= \sum_n P_\eta(k|n) p_n(g) \\ &= 
	\sum_n P_\eta(k|n) \sum_N \pi_N p^N_n(g) 
	\nonumber
	~.
\end{align}
One can reconstruct the entire probability $p_n(g)$ by ``inverting'' the first equation,
as described in \cite{lee2004towards,achilles2004photon}, and then finding $\pi_N$ 
and $p^N_n(g)$ by conditioning. The resulting reconstruction protocol would then be equivalent 
to the one presented  in Sec.~\ref{s:imperfectsources}

\section{Remark about numerical experiments}
The numerical experiments are performed by generating $R$ Haar random unitary matrices $g$, 
calculating the superoperator $\mathcal L_{\nu' g,\alpha\beta} = 
\bra{\nu'} U(g)^\dagger \ket\alpha\bra\beta U(g)\ket{\nu'} $, and then finally studying 
the rank of the Gramian matrix $\mathcal L^\dagger \mathcal L$. Complete tomography is possible 
when such Gramian matrix has full-rank. 
We have observed that numerically generated Haar random matrices were always capable 
of producing a full rank Gramian when $R\ge R_{N,M}$ unitary matrices are employed. 
This numerically shows that our lower bound is achievable with Haar random configurations, 
at least for the sizes tested  ($M\le 6$ and $N\le 10$). 

Moreover, we have also numerically generated random linear optics configurations 
following the universal construction of Ref.~\cite{clements2016optimal}. We have observed 
that even when we consider random matrices obtained with random phase shifters and 
beam splitters (i.e. phase shifters and transmissivities 
sampled from independent uniform distributions), we obtain the same results. 
Such random matrices are not 
necessarily Haar random, but still they are ``random enough'' to provide the necessary 
information for complete state reconstruction. A more sophisticated construction can be 
obtained following the techniques of Ref.~\cite{russell2017direct}. 
On the other hand, random phase shifts 
and fixed transmissivities are sufficient for $M=2$, as also shown by the Newton and 
Young algorithm \cite{newton1968measurability}, but not for $M > 2$. 
Therefore, we conjecture that linear optics configurations with uniformly random 
phase shifters and trasnmissivities are sufficient for complete tomography with 
minimal measurement settings. 

\section{Numerical complexity of state reconstruction}
We study the numerical complexity of a simple state reconstruction protocol, based 
just on the maximum likelihood estimator 
\begin{equation}
\rho_{\rm MLE} := 
(\mathcal L^\dagger \mathcal L)^{-1} \mathcal L^\dag[p]~,
\end{equation}
where $\mathcal L_{\nu' g,\alpha\beta} = 
\bra{\nu'} U(g)^\dagger \ket\alpha\bra\beta U(g)\ket{\nu'} $, and 
$p_{\nu',g} = \bra{\nu'} U(g)^\dagger \rho U(g)\ket{\nu'}$ are the measured 
probabilities. Here we do not study any possible simplification in the evaluation 
of matrix permanents and consider the cost of evaluating the independent permanents 
$\bra\beta U(g)\ket{\nu'}$, without any other assumption. 
Indeed, as discussed in the main text, $\bra\beta U(g)\ket{\nu'}$ can be written as the 
permanent of a $N\times N$ matrix. Using Ryser's formula \cite{ryser1963combinatorial} 
the cost of evaluating such permanent is $\mathcal O(2^{N-1}N)$ arithmetic operations. 
Such cost is independent
on the number of input and output modes $M$ and $M'$. This estimate can be highly improved, 
for instance it is known that when there are many 
collisions (e.g. $M>N$) the permanent is easier \cite{aaronson2011computational}. However, here 
we do not consider any simplifying assumption and study just the numerical complexity with
Ryser's algorithm. In that case, one has to evaluate 
\begin{equation}
	\mathcal O(R_{N,M,M'}D_{N,M}D_{N,M'})  \approx 
	\mathcal O\left(R_{N,M}\frac{D_{N,M-1}}{D_{N,M'-1}}D_{N,M}D_{N,M'}\right)  
\end{equation}
of such permanents.

When $N\gg M, M'$ then  $R_{N,M}\approx D_{N,M} \approx N^{M-1}$  so the complexity of 
building $\bra\beta U(g)\ket{\nu'}$  is $\mathcal O(N^{3M-2} 2^{N-1})$. The maximum likelihood
estimator then requires a polynomial number of operations in matrix dimensions, so 
the total numerical complexity is ${\rm poly}(N^{2M-2},2^{N-1}) \simeq  
{\rm poly}(D_{N,M}, N 2^{N-1})$, which is polynomial in terms of the Hilbert space dimension, 
but exponential in terms of number of photons. The regime where 
$R_{N,M,M'}\approx 1$ reduces the experimental complexity, at least in terms of number of 
configurations, but does not reduce the numerical overhead:  as 
$M'\approx 2M-1$  (Eq.6 from the main text), the complexity is still 
$\mathcal O(N^{M-1}N^{M'-1}N 2^{N-1}) \approx \mathcal O(N^{3M-2} 2^{N-1})$. 
Nonetheless, 
this overhead can possibly be highly improved by exploiting numerical simplifications that
takes into account particle collisions \cite{aaronson2011computational}, which are 
inevitable when $N\gg M$. 

On the other hand, in the opposite regime $M\gg N$ particle collisions are unlikely.
In this regime $R_{N,M} \approx D_{N,M+1} \approx M^N/N!$ and we get 
a complexity ${\rm poly}((M^{N}/N!)^3, N 2^{N-1}) 
$.  
Therefore, in both limits we obtain the scaling $
\mathcal O({\rm poly}(D_{N,M}, N 2^{N-1})).
$

\end{document}